\documentclass[pageno]{jpaper}

\usepackage{amsmath}

\usepackage[normalem]{ulem}

\begin{document}

\title{An Architecture for Improved Surface Code Connectivity in Neutral Atoms}
\date{}
\author{
    Joshua Viszlai$^{1}$, Sophia Fuhui Lin$^{1}$, Siddharth Dangwal$^{1}$, Jonathan M. Baker$^{2, 3}$, Frederic T. Chong$^{1}$\\
    \normalsize{$^{1}$University of Chicago} \\
    \normalsize{$^{2}$Duke Quantum Center} \\
    \normalsize{$^{3}$University of Texas, Austin}
}

\maketitle
\thispagestyle{empty}


\begin{abstract}

  In order to achieve error rates necessary for advantageous quantum algorithms, Quantum Error Correction (QEC) will need to be employed, improving logical qubit fidelity beyond what can be achieved physically. 
  As today's devices begin to scale, co-designing architectures for QEC with the underlying hardware will be necessary to reduce the daunting overheads and accelerate the realization of practical quantum computing.
  
  In this work, we focus on logical computation in QEC. We address quantum computers made from neutral atom arrays to design a surface code architecture that translates the hardware's higher physical connectivity into a higher logical connectivity. We propose groups of interleaved logical qubits, gaining all-to-all connectivity within the group via efficient transversal CNOT gates. Compared to standard lattice surgery operations, this reduces both the overall qubit footprint and execution time, lowering the spacetime overhead needed for small-scale QEC circuits.

  We also explore the architecture's scalability. We look at using physical atom movement schemes and propose interleaved lattice surgery which allows an all-to-all connectivity between qubits in adjacent interleaved groups, creating a higher connectivity routing space for large-scale circuits. Using numerical simulations, we evaluate the total routing time of interleaved lattice surgery and atom movement for various circuit sizes. We identify a cross-over point defining intermediate-scale circuits where atom movement is best and large-scale circuits where interleaved lattice surgery is best. We use this to motivate a hybrid approach as devices continue to scale, with the choice of operation depending on the routing distance.

\end{abstract}

\begin{figure}
    \centering
    \includegraphics[width=\linewidth]{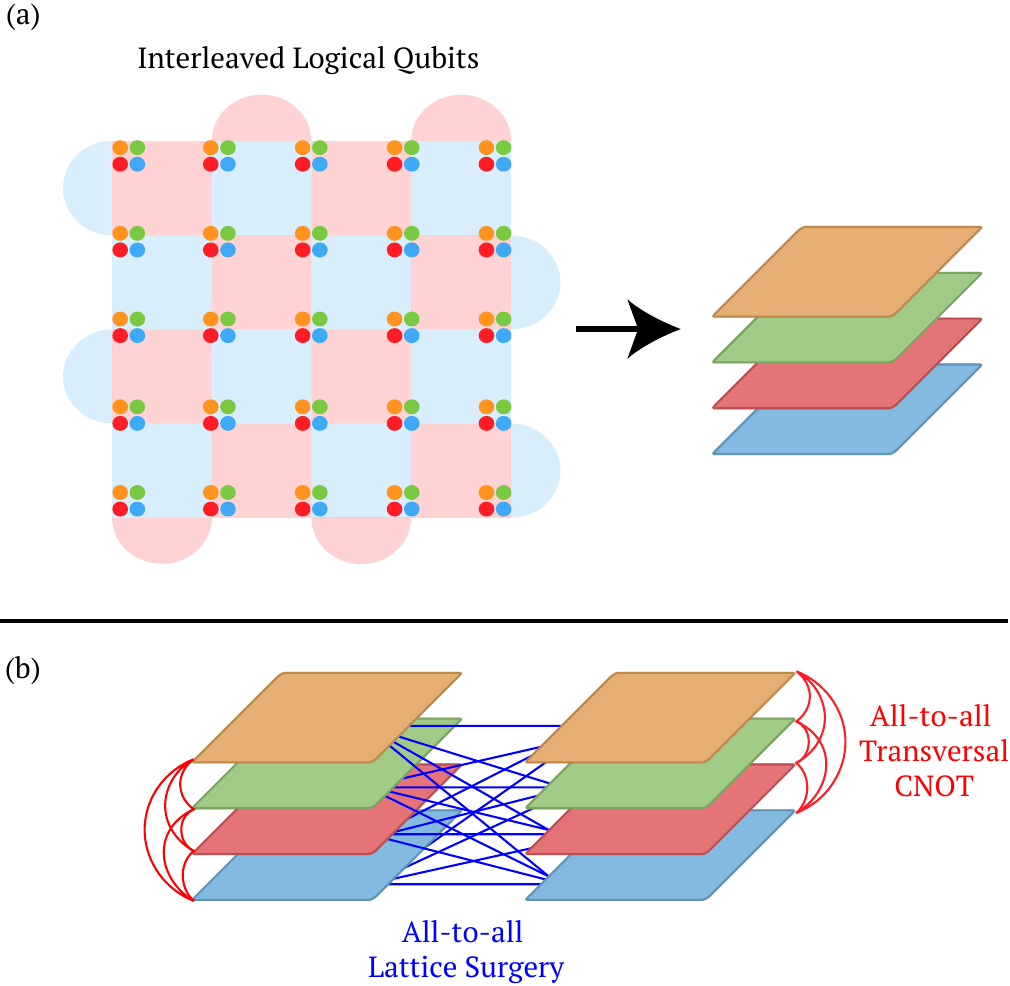}
    \caption{High-level overview of our proposed architecture which is further described in Figure~\ref{fig:multi-qubit}. (a) An interleaving of 4 surface code qubits into one group, visualized as a stack. (b) The enabled all-to-all connectivity within a group via a fast, transversal CNOT and between adjacent groups via interleaved lattice surgery.}
    \label{fig:overview}
\end{figure}

\section{Introduction}

Today, quantum computers exist in the Noisy Intermediate-Scale Quantum (NISQ) era. Devices of this era have 10s to 100s of qubits with relatively high error rates. Although there's been promising research to realize practical quantum advantage with algorithms tailored to today's NISQ devices, many algorithms with known quantum advantage require error rates much lower than can be achieved in NISQ~\cite{shor1999polynomial,grover1996fast}. 
It's believed that running these disruptive algorithms will have to wait until the era of Fault-Tolerant Quantum Computing (FTQC), where each logical qubit is encoded in 100s of physical qubits via some Quantum Error Correcting (QEC) code. By increasing the number of physical qubits, we exponentially decreases the logical qubit's error rate, allowing us to reach the low error rates needed for advantageous algorithms.

The cost of fault-tolerance is quite high, and comes with its own set of unique challenges. 
Problems addressed for NISQ machines such as compiling to hardware gate sets, mapping program qubits to physical qubits, and routing multi-qubit operations will all be different at this new, logical level. This motivates new work to address these challenges and trade-offs in the FTQC regime.

Furthermore, the underlying hardware of quantum computers is highly varied. Many proposed methodologies for creating qubits exist including superconducting~\cite{krantz2019quantum}, trapped ions~\cite{bruzewicz2019trapped}, nitrogen-vacancy centers~\cite{schirhagl2014nitrogen}, and neutral atom arrays~\cite{singh2022dual, bluvstein2022quantum}. This diversity in hardware implementations motivates a tailored approach to error correction. By co-designing QEC architectures with the hardware, we can better utilize their strengths and inform where experimental progress should focus its efforts for the FTQC regime.  

In this work we design and evaluate a QEC architecture for neutral atom arrays using the surface code, shown in Figure~\ref{fig:overview}. 
Our architecture focuses on improving multi-qubit logical operations in the surface code. We translate neutral atom array's higher physical connectivity to a higher logical connectivity, reducing the spacetime costs of surface codes compared to a standard, hardware-agnostic architecture. Specifically, we propose a device tiled by groups of \textbf{interleaved} qubits. We discuss the hardware costs to implement this architecture based on recent devices, and analyze the performance as device size scales. At small-scales within a single group, our architecture enables a fast, transversal CNOT and an all-to-all connectivity, removing the need for routing ancilla and reducing execution time compared to lattice surgery. For larger scales that would support advantageous quantum algorithms, we compare routing via mid-circuit atom movement and a proposed implementation of lattice surgery between interleaved groups that allows all-to-all connectivity for logical qubits in adjacent groups. Using compute time as our figure of merit and assuming a fixed qubit budget, we identify a crossover point of $\sim$50 program qubits, below which routing via atom movement is best, and above which routing via lattice surgery is best. We use these results to motivate a hybrid approach as devices continue to scale.

To summarize, we present the following contributions:

\begin{itemize}
    \item A surface code architecture for neutral atom arrays where logical qubits are interleaved. For $\sim$4-16 logical qubits, this provides an all-to-all connectivity via a transversal CNOT. This removes the need for qubits serving as routing ancilla, and we find the transversal CNOT has a lower logical error rate, further reducing the costs of near-term QEC.

    \item A modified implementation of lattice surgery to address large-scale systems that permits an all-to-all connectivity between adjacent groups of interleaved qubits. This produces a higher connectivity routing space, reducing routing costs for large-scale surface code computations.
    
    \item An analysis of the overall compute time for running future Clifford + T circuits in our architecture. We compare two routing strategies and analyze the sensitivity of device-level features.
    

    \item Simulation methodologies for evaluating the logical error rate and routing time overhead for surface codes in neutral atom arrays.
\end{itemize}

\section{Background}
For background on quantum computing fundamentals, we refer to~\cite{ding2020quantum, nielsen2001quantum, rieffel2011quantum}. Here, we provide relevant background for Quantum Error Correction (QEC) and neutral atom arrays.

\subsection{Quantum Error Correction}
In order to execute most quantum algorithms, we need orders of magnitude lower error rates than we can achieve physically. QEC is the most promising path towards achieving these requisite error rates. A typical QEC scheme entails encoding \textbf{logical qubits} over many physical qubits. For example, the quantum analog of the classical repetition code can encode a single logical qubit as follows
\begin{align*}
    |0 \rangle_L = |000\rangle \\
    |1 \rangle_L = |111\rangle
\end{align*}
In this example, a logical qubit is encoded in 3 physical qubits with the hamming distance between these encodings, often called the \textbf{code distance} being 3. If we think of an error as a bit flip, the code distance describes how many physical errors are needed to create a logical error. By increasing the number of physical qubits, we increase the code distance and improve our resilience to physical errors. This is the key idea of QEC: we trade physical resources, qubits, for lower logical error rates. If our physical error rate is below some threshold, then this trade-off is heavily one-sided~\cite{aharonov1997fault,kitaev2003fault,knill1998resilient}. By increasing our code distance linearly we exponentially decrease our logical error rate.

We also need to repeatedly find and correct physical errors to minimize the probability that they grow to a logical error. Since we can't measure our qubits without destroying our quantum information, many QEC codes use \textbf{stabilizers}~\cite{gottesman1997stabilizer,poulin2005stabilizer}. A code's stabilizers are a set of commuting observables that define a shared eigenspace where the logical qubit lives. A physical error may cause the logical qubit to deviate outside this eigenspace, but measuring the stabilizers projects the qubit back into the eigenspace. This results in a discretization of physical errors. Physical errors become bit-flips or phase-flips of physical qubits, and a change in the eigenvalue measured for a given stabilizer, called a \textbf{syndrome}, indicates such an error on one or more physical qubits. The stabilizers implicitly perform \textbf{parity checks} between sets of data qubits, which reveals enough partial information to correct errors without destroying our quantum states. For example, in the above repetition code our stabilizers are: $ZZI, IZZ$, i.e. parity checks between qubits 1,2 and 2,3. If we initially measure $+1$ eigenvalues for both, then we know all 3 physical qubits have the same bit value, but if later we measure the stabilizers and get $-1$ for both, then we know a physical error has occurred. Specifically, the most likely error is the middle qubit had a bit-flip, disrupting both checks. This process of deducing the most likely physical error from syndromes is called \textbf{decoding}. 

In practice, QEC codes need to detect and correct bit-flips and phase-flips. Additionally, we need to be able to perform gates on the logical qubits while preserving our resilience to physical errors. For more background on QEC and stabilizer codes, see~\cite{fowler2012towards,gottesman1997stabilizer}.

\subsubsection{Surface Code}
\begin{figure}
    \centering
    \includegraphics[width=0.5\linewidth]{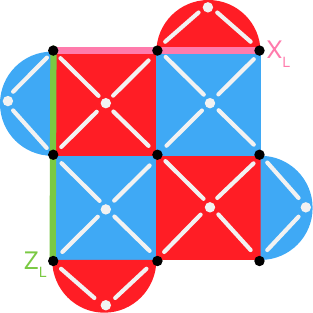}
    \caption{A distance 3 surface code. Z (X) stabilizers are denoted by red (blue) faces, and data qubits are denoted by black vertices. Stabilizers are a product of Z (X) of all data qubits incident to the face. A white vertex indicates an ancilla qubit used for measuring a stabilizer via 2-qubit gates on white edges. Green (pink) chains indicate a product of Z (X) on data qubits that corresponds to the logical Z (X) observable.}
    \label{fig:surface_code}
\end{figure}


The Surface Code is a QEC code embedded on a 2D nearest-neighbor grid~\cite{horsman2012surface,fowler2012towards}. Stabilizers are either X-type: $\{XXXX$, $XX\}$, or Z-type: $\{ZZZZ$, $ZZ\}$, although variants like the $XZZX$ surface code exist~\cite{bonilla2021xzzx}. The surface code's local connectivity means it can be implemented on a device with 2D nearest neighbor connectivity, such as superconducting devices~\cite{google2023suppressing}. It also has a high threshold of $\sim 0.1\%$, meaning if the physical error rate is below $0.1\%$, increasing the code distance will exponentially decrease the logical error rate. These two features: low connectivity requirement and high threshold, have made the surface code a popular QEC code for early demonstrations of FTQC~\cite{google2023suppressing,krinner2022realizing}. Figure~\ref{fig:surface_code} shows a surface code in the rotated picture, where the 2D grid is rotated $45\degree$. 

\subsubsection{Operations in the Surface Code}

\begin{figure}
    \centering
    \includegraphics[width=0.4\linewidth]{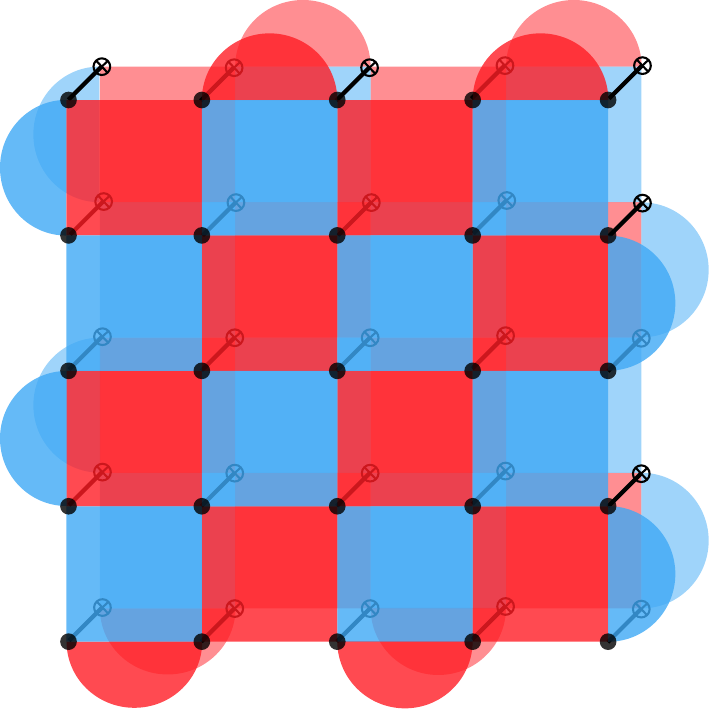}
    \caption{A transversal CNOT between two distance 5 surface code qubits.}
    \label{fig:t_cnot}
\end{figure}

Measuring the surface code's stabilizers is sufficient for memory operations on the logical qubit, however, for FTQC we need to be able to perform universal quantum computation on logical qubits. This requires a fault-tolerant gate set, where we have logical gates that are also protected by our QEC code. In general, the set of fault-tolerant gates for the surface code is the Clifford group: $\{X,Y,Z,H,S,CNOT\}$~\cite{horsman2012surface,fowler2018low}. 
To achieve universality, however, we also need a non-Clifford gate such as the $T$ gate. Since these are not protected by the surface code, they require more involved techniques to reach requisite logical error rates such as T state distillation~\cite{bravyi2012magic, litinski2019magic} or post-selected state preparation~\cite{choi2023fault}.

For multi-qubit interactions, a simple and fast operation is to perform CNOT transversally, as shown in Figure~\ref{fig:t_cnot}. However, this imposes a high connectivity requirement which a nearest neighbor connectivity cannot support. For many platforms, like superconducting devices, this isn't feasible. As a result, alternative methods for performing logical operations between surface code qubits have been designed~\cite{horsman2012surface}. 

\begin{figure}
    \centering
    \includegraphics[width=0.75\linewidth]{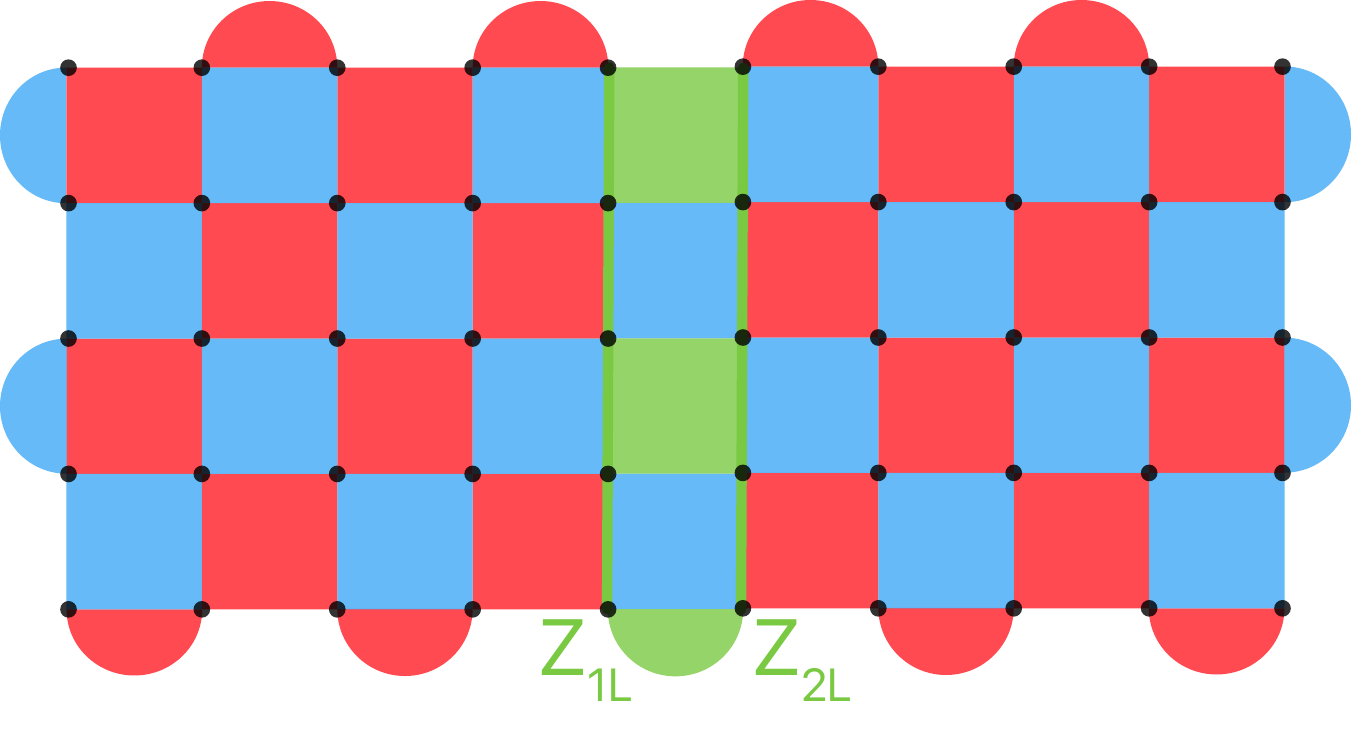}
    \caption{A lattice surgery operation where two qubits are \textbf{merged} by measuring new Z stabilizers (green). The product of these new stabilizers is equivalent to $Z_{1L} Z_{2L}$, meaning the merge accomplishes a logical $ZZ$ measurement.}
    \label{fig:ls_merge}
\end{figure}

\begin{figure}
    \centering
    \includegraphics[width=\linewidth]{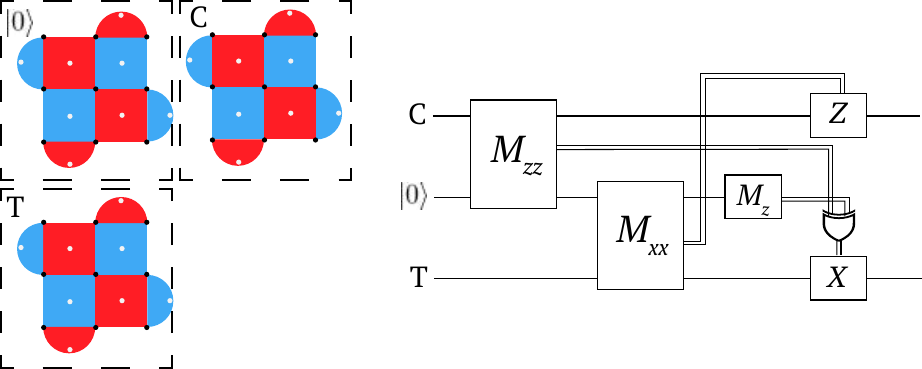}
    \caption{A measurement based CNOT gate that can be achieved using two lattice surgery operations at the cost of an extra logical qubit as an ancilla. (Left) Physical layout of logical qubits, including ancilla. (Right) Circuit-level description of the CNOT. $M_{ZZ}$ and $M_{XX}$ are performed using lattice surgery as described in Figure~\ref{fig:ls_merge}.}
    \label{fig:lattice_cnot}
\end{figure}

\textbf{Lattice Surgery} is a popular option for performing gates, including multi-qubit gates, on devices with only 2D nearest neighbor connectivity. In lattice surgery, we merge and split surface code patches to accomplish Pauli-product measurements. Figure~\ref{fig:ls_merge} shows an example of a lattice surgery operation on two qubits that accomplishes a $ZZ$ measurement. To realize a CNOT between two qubits, we can use a measurement-based protocol that consists of a $ZZ$ and $XX$ measurement~\cite{horsman2012surface} as shown in Figure~\ref{fig:lattice_cnot}. Alternatively, prior work has shown how to describe circuits as only Pauli-product measurements, removing the need to synthesize other gates~\cite{litinski2019game}.

It's important to note that since the result of the Pauli-product measurement is revealed via new stabilizers which may have \textbf{measurement errors}, we need to repeat the stabilizer measurement to make sure the result is correct. Specifically, if our code distance is $d$, we need to perform $d$ measurements of the new stabilizers. 

\subsection{Neutral Atom Arrays}\label{sec:atom_arrays}
Neutral atom quantum computing is a newer QC platform that relies on arrays of atoms trapped in rearrangeable optical tweezers~\cite{kaufman2012cooling,lester2015rapid,muldoon2012control}. Although up to 3D arrays are technically possible, for this work we assume a 2D array which is more feasible when supporting digital gates on atoms. The qubit states, $|0\rangle$, $|1\rangle$, are defined as two hyperfine energy levels of the atom, which have long coherence times on the order of $\sim100$ms - $1$s. Single qubit gates are performed by lasers that drive Raman transitions between these two states, which typically take $\sim1 \mu$s~\cite{singh2022dual, bluvstein2022quantum}. 

\subsubsection{Rydberg Gate}\label{sec:rydberg}
Multi-qubit gates between atoms are traditionally performed via the Rydberg gate. Atoms are coupled to a highly excited Rydberg state which induces a van der Waals interaction with nearby atoms~\cite{saffman2016quantum}. This allows for a native gate between any number of atoms in some interaction radius which takes $\sim5 \mu$s. More specifically, the interaction can be used to realize any $C^nZ^k$ gate--a controlled-phase gate with $n$ controls and $k$ targets. Since the interaction radius can extend past multiple atoms, this allows for a larger connectivity than nearest-neighbor. An important caveat is two Rydberg gates cannot be executed in parallel if their interaction zones overlap. 


\section{Problem and Motivation}

Although the theory of QEC has existed for many years, the implementation of QEC codes on real hardware is still in its infancy, with only a few experimental demonstrations. Furthermore, given the highly varied landscape of qubit technologies, efficient realization of QEC codes will be distinct between devices. Given the necessity of QEC for most quantum applications, evaluating the relative capability of hardware platforms therefore depends on their efficacy in implementing QEC codes, or their \textbf{logical} performance. In this work we aim to improve the logical performance of a single hardware platform, neutral atom arrays, for a given QEC code, the surface code. Our goal is that such work will not only improve the logical performance, but also inform experimentalists of which device features are most important for FTQC.

\subsection{Hardware Platforms}
    Among the various hardware platforms, superconducting devices have seen lots of popularity, with recent experimental demonstrations of a single logical qubit at the surface code threshold~\cite{google2023suppressing,krinner2022realizing}. These devices are fast and can scale in the number of qubits. However, short coherence times and fabrication defects~\cite{brink2018device,devoret2004superconducting} present challenges for running large-scale computations needed for FTQC. Connectivity is also limited, with typical devices having at most a 2D grid with nearest-neighbor connectivity. 
    
    Neutral atom arrays are newer platform, with inherent strengths that are attractive for FTQC. Atoms are homogeneous, and so do not suffer from fabrication defects. Scaling to more qubits is also relatively easy as experimental demonstrations of more than 250 atoms have been achieved~\cite{ebadi2021quantum}.
    Additionally, primitive entangling operations can reach beyond nearest neighbors~\cite{saffman2016quantum} and atoms can be coherently transported during computation~\cite{bluvstein2022quantum}, allowing for configurable, arbitrary geometries. These enable the atoms to support a high physical connectivity, motivating its translation to a high logical connectivity. 
    Despite these strengths, neutral atom devices still have difficulties. Gate fidelities are low and measurement schemes are slow and global, inhibiting the ability to do QEC which requires repeated measurement of ancilla. Recent work however, has shown notable improvement in gate fidelities~\cite{evered2023high}, and experimental demonstrations of dual-species atom arrays presents a way to measure ancilla without disturbing data qubits, enabling QEC~\cite{singh2022dual}. As such, we expect these devices to continue to improve in the coming years, and believe they are a promising platform for QEC.

\section{Related Work}

A related architectural study on co-designing QEC and hardware has proposed a surface code architecture tailored to superconducting qubits with cavities~\cite{duckering2020virtualized}. 
While this work makes a strong first step towards hardware tailored QEC, it does not evaluate the scalability of the architecture beyond providing a lattice surgery primitive. 
The use of cavities also prohibits parallel execution of gates on multiple pairs of qubits in the same cavity, serializing both stabilizer measurements and execution of logical gates. In our work, there is no virtualization of logical qubits. So while we have similar logical organization of qubits, we don't pay the same serialization penalty. We execute all stabilizers in the same measurement cycle and enable concurrent lattice surgery operations using high capacity ancilla channels. We also study the scalability of our proposed architecture which benefits from the neutral atom hardware. 



Prior systems studies on neutral atoms~\cite{baker2021exploiting, patel2022geyser, nottingham2023decomposing} also exist, but focus on NISQ-era applications. In our work we instead address QEC applications, where the circuits that need to be executed are largely limited to stabilizer circuits. We address this smaller space of possible circuits while also considering recent hardware constraints and capabilities including: limited local control, dual species interactions, and mid-circuit atom movement.

Prior theoretical work on fault-tolerant architectures with neutral atoms~\cite{cong2022hardware} studies engineering biased noise and bias-preserving gates but is limited to small QEC codes. This has potential applications within our study of surface codes, which we briefly discuss in Section~\ref{sec:circ_level}.

Prior experimental work on surface codes with neutral atoms~\cite{bluvstein2022quantum} studies execution of a single stabilizer round using movement operations and local Rydberg gates, providing good motivation for the realization of QEC in neutral atom devices. The authors find atom loss is the dominant error source during movement, but can be mitigated by lowering the movement speed. Given the high number of stabilizer circuits necessary for running large-scale QEC applications, however, movement-based stabilizer circuits may incur large time overheads. In light of this, we explore stabilizer circuits without movement, retaining high connectivity through longer range interactions and limiting movement for use in routing operations at the logical level, which occur less frequently than stabilizer circuits.

\section{Interleaved Qubits in Atom Arrays}

\begin{figure}
    \centering
    \includegraphics[width=\linewidth]{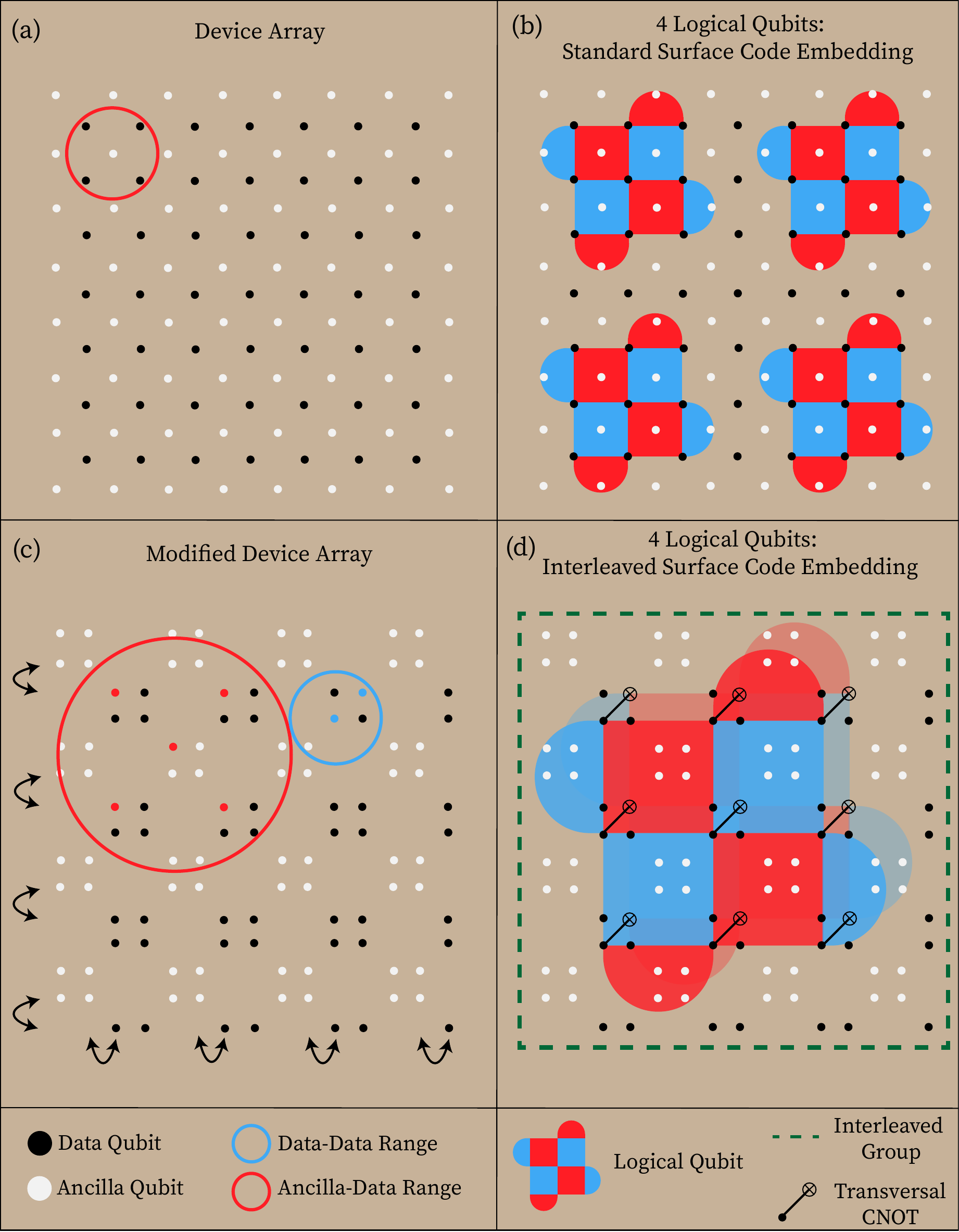}
    \caption{Overview of differences between a standard surface code embedding and the proposed interleaved embedding. (a) Neutral atom array for the standard surface code embedding. The Rydberg interaction range (red) only needs to be large enough to perform two qubit gates between nearby ancilla and data. The data-data interaction rage is omitted since it is unused in the standard surface code embedding. (b) Software level assignment of physical qubits to logical qubits.
    A logical CNOT can be accomplished using lattice surgery as described in Figure~\ref{fig:lattice_cnot}. (c) Modified device array to support the proposed interleaved embedding. The ancilla-data range is increased to support the stabilizer measurements and the data-data range enables a logical CNOT between logical qubits. (d) Software level assignment of physical qubits to logical qubits in the interleaved embedding. Each of the 4 logical qubits occupies a position (top left, top right, bottom left, bottom right) in the qubit clusters. For clarity, the surface code overlay is only shown for two of the 4 logical qubits. An example transversal CNOT is shown between the bottom left and top right logical qubits.} 
    \label{fig:multi-qubit}
\end{figure}

We target quantum computers made from neutral atom arrays which can support a higher than nearest-neighbor connectivity as discussed in Section~\ref{sec:rydberg}. We propose an architecture which is a 2D array of \textbf{groups} of many interleaved surface code qubits to translate this high physical connectivity into a high logical connectivity, leading to a reduced time and space overhead for fault-tolerant logical operations. In Section~\ref{sec:single-iq} we describe a single interleaved group and discuss differences compared to a standard architecture on few qubits. Then in Section~\ref{sec:multi-iq} we look at the scalability of the architecture to 100s of logical qubits. We discuss the choice of using lattice surgery or atom movement between interleaved groups and analyze the associated time and space overheads. Finally we describe our simulation methodology and present numerical results evaluating the architecture in Section~\ref{sec:evaluate}.


\begin{figure}
    \centering
    \includegraphics[width=\linewidth]{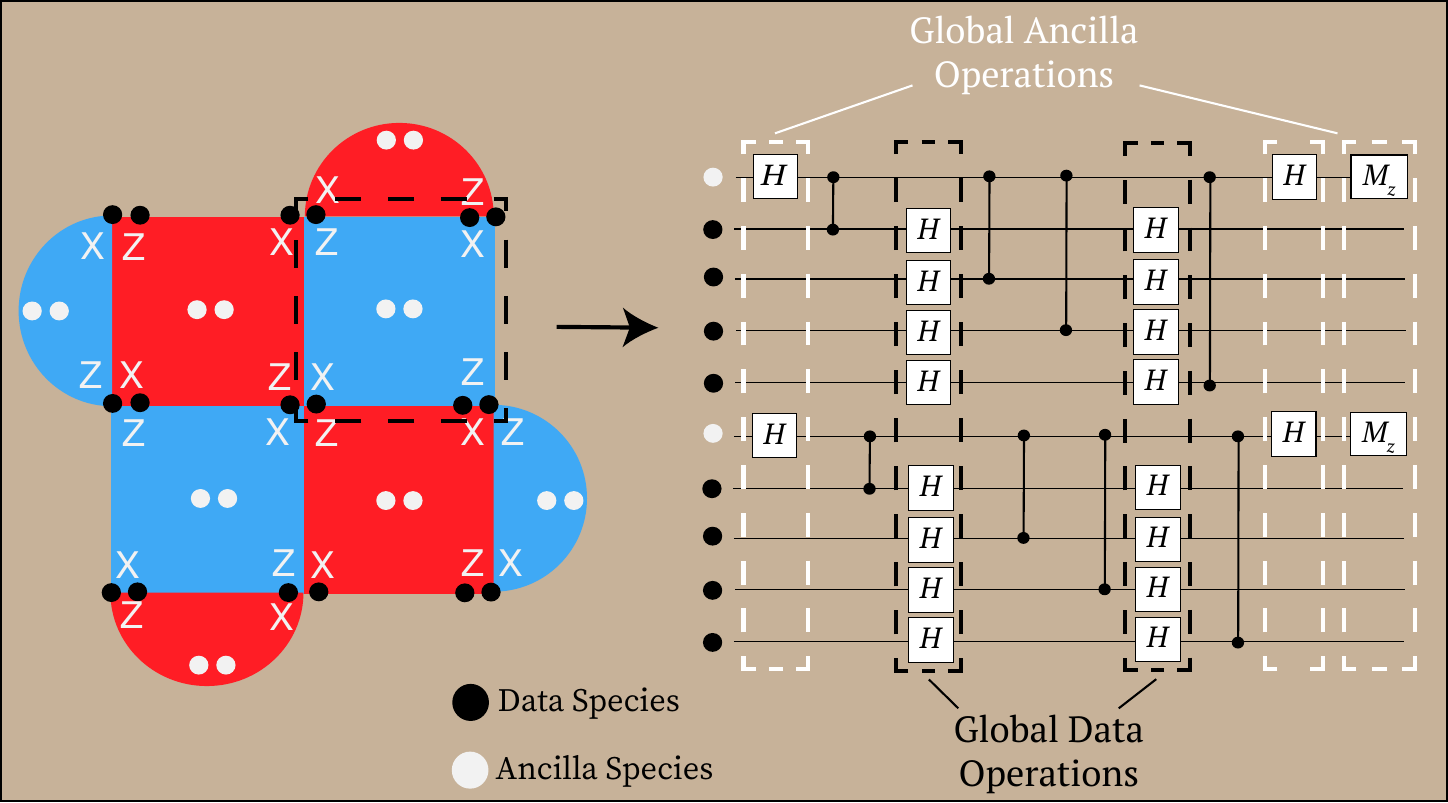}
    \caption{Stabilizer circuit structure of two interleaved stabilizers using the ZXXZ surface code. CZ gates between data and ancilla are serialized when interaction radii intersect. Using the ZXXZ surface code all stabilizers are identical, allowing H gates to be globally aligned per species. This improves compatibility with existing hardware where only Rz and CZ gates are local, and X-Y rotations are global.}
    \label{fig:stab_circ}
\end{figure}

\subsection{Interleaved Surface Code Qubits}\label{sec:single-iq}

Figure~\ref{fig:multi-qubit} depicts a single interleaved group of surface code qubits in our architecture. In this section we further detail this in three parts: the hardware level, the software level, and the circuit level.

\subsubsection{Hardware Level}

In our proposal, compared to a standard surface code embedding, the initial atom array is rearranged to create clusters of data and ancilla. The example shown in Figure~\ref{fig:multi-qubit}c creates clusters of 4 physical qubits to interleave 4 logical qubits, which we call an \textbf{interleaved group}. In general an interleaved group of $n$ logical qubits requires a device array with clusters of $n$ physical qubits. Since arbitrary 2D geometries have been generated in these arrays experimentally, this is no more difficult than an array for a standard surface code embedding. Additionally, to enable a transversal CNOT between logical qubits, the interaction range between data qubits only needs to span a cluster of physical qubits (4 qubits in the example shown in Figure~\ref{fig:multi-qubit}c) which is comparable to the standard embedding. Two notable requirements, however, are: \raisebox{.5pt}{\textcircled{\raisebox{-.9pt} {1}}} two qubit gates between ancilla and data qubits require larger interaction ranges, and \raisebox{.5pt}{\textcircled{\raisebox{-.9pt} {2}}} measuring QEC stabilizers requires measuring ancilla qubits without measuring data qubits. Typical neutral atom arrays using atoms of the same species have practical limits on the radius of the Rydberg gate due to achievable Rydberg states~\cite{saffman2016quantum, evered2023high}, impeding \raisebox{.5pt}{\textcircled{\raisebox{-.9pt} {1}}}, and global measurement, preventing \raisebox{.5pt}{\textcircled{\raisebox{-.9pt} {2}}}. Recent work, however, has experimentally demonstrated a dual-species atom array~\cite{singh2022dual,sheng2022defect}. In these devices, each atom species can be measured separately. This is still a global operation per species, but it means we can assign one species to data qubits, one species to ancilla qubits, and then measure our ancilla species without disturbing our data qubits, solving requirement \raisebox{.5pt}{\textcircled{\raisebox{-.9pt} {2}}}. Furthermore, interactions between atoms of different species can be longer than between atoms of the same species by using Förster resonances~\cite{beterov2015rydberg}. As a result, we can have notably larger interaction radii between ancilla and data qubits, solving requirement \raisebox{.5pt}{\textcircled{\raisebox{-.9pt} {1}}}.

\subsubsection{Software Level}
At the software level, we modify the assignment of physical qubits to logical qubits compared to the standard surface code embedding. Given the new device array, we assign each logical qubit in the interleaved group a position. For the 4 logical qubit example in Figure~\ref{fig:multi-qubit}, the positions are the top left, top right, bottom left, and bottom right of each physical qubit cluster. Each logical qubit is comprised of the physical qubits at its position in each cluster. 
The immediate benefit of this encoding is the ability to perform a transversal CNOT between logical qubits. This circumvents the need for lattice surgery between logical qubits in the same interleaved group, which requires additional logical qubits to serve as ancilla, as shown in Figure~\ref{fig:lattice_cnot}.

\subsubsection{Circuit Level}\label{sec:circ_level}

QEC circuits consist
almost entirely of repeated stabilizer measurement circuits.
The execution of this circuit therefore defines an atomic unit, analogous to a clock cycle in classical computing. Figure~\ref{fig:stab_circ} shows an example stabilizer circuit structure. Here, we discuss two hardware features of this circuit: \raisebox{.5pt}{\textcircled{\raisebox{-.9pt} {1}}} Since existing neutral atom devices have limited local addressability~\cite{nottingham2023decomposing}, the use of the ZXXZ surface code~\cite{bonilla2021xzzx} mitigates qubit control costs by allowing for globally aligned H gates per species. The ZXXZ variant is locally equivalent to the standard surface code and although we focus on its use to improve compatibility with existing hardware, it's known to have improved performance in the presence of biased noise, which recent work shows can be engineered in neutral atom arrays~\cite{cong2022hardware}. The ZXXZ surface code has also been implemented on superconducting devices~\cite{google2023suppressing}, further motivating its viability.
\raisebox{.5pt}{\textcircled{\raisebox{-.9pt} {2}}} In the interleaved architecture, CZ interactions in stabilizer circuits are mediated by interactions between data and ancilla atoms within some radius. Due to the interleaving, the desired interaction radii for each of the logical qubits intersect. As a result, we must serialize CZ gates to an extent.

\paragraph{\textup{\textbf{Cost of Serialization:}}}
For an interleaved group of $k$ qubits, serialization of CZ gates leads to a $k\times$ increase in the stabilizer circuit depth. For neutral atom devices, however, this is not a large cost to pay. Atoms have long coherence times compared to gate times and the increase in depth does not add any new gate operations, just idle time for ancilla qubits. As such, we do not expect a notable increase in error rates. The additional \textbf{compute time} might still be of concern, however, we note measurement operations take $\sim 10$ms whereas gate operations take $\sim1$-$5\mu$s~\cite{singh2022dual, bluvstein2022quantum}. This means the length of a stabilizer measurement cycle is dominated by the measurement time, not the gate time, so serialization of gate operations will have a negligible impact on the overall execution time.

\subsection{Logical Multi-Qubit Gates}
 We evaluate two methods for multi-qubit gates in this work: a transversal CNOT and a lattice surgery CNOT. Both of these can be implemented in a dual-species atom array. In our interleaved architecture, a transversal CNOT becomes possible as described in 
Figure~\ref{fig:multi-qubit}d. In a regular surface code architecture, a lattice surgery CNOT can be accomplished using the circuit in Figure~\ref{fig:lattice_cnot}, but with a slight modification. The initial issue with lattice surgery in the dual-species arrays is we need to measure logical ancilla, which requires measuring the ancilla's underlying data qubits. However, measurement is global per species and so attempting to measure the data species would measure all of our logical qubits, collapsing our circuit. To circumvent this issue, instead of measuring the underlying data atoms we need to swap the information from the data atoms to the ancilla atoms. 
We then measure the ancilla as normal and treat the measurement results as if it were data qubits. Again due to global measurement, this operation needs to be aligned with the stabilizer circuits occurring on other qubits. 

\paragraph{\textup{\textbf{Transversal vs Lattice Surgery:}}}
To compare the efficacy of a CNOT performed transversally and via lattice surgery, we can look at the spacetime overhead of each. Considering the space costs, the all-to-all connectivity within a single interleaved group requires no logical ancilla for transversal CNOTs and so there is no additional space overhead. A lattice surgery CNOT, however, requires an additional logical ancilla and further logical ancilla for routing. In total this can increase the qubit count by $1.5x-2x$ depending on the routing layout. Looking at the time costs, a transversal CNOT is a single operation, with a conservative estimate requiring $d$ stabilizer rounds. Performing a lattice surgery CNOT requires two operations, each of which is $d$ stabilizer rounds, totaling $2d$ measurement rounds. Therefore a lattice surgery CNOT required in the standard surface code embedding requires $1.5x-2x$ more physical qubits and takes $2x$ more measurement rounds to complete compared to a transversal CNOT enabled by interleaved logical qubits.

\subsection{Operations Between Interleaved Groups}\label{sec:multi-iq}

In practice, the achievable interaction radius has a ceiling. If we just keep adding qubits to a single interleaved group as our device grows, eventually the ancilla will not be in range of the data qubits. Additionally, the serialization of operations will at some point dominate the compute time. The consequence is a single interleaved group of qubits is not scalable. 

To scale to larger devices, we propose multiple groups with a fixed set of interleaved qubits per group. This comes with a new problem, however, as operations between qubits that are not in the same group cannot immediately be performed transversally using the Rydberg gate. In this section, we discuss two possible methodologies to address this.

\subsubsection{Interleaved Lattice Surgery}

\begin{figure}
    \centering
    \includegraphics[width=0.75\linewidth]
    {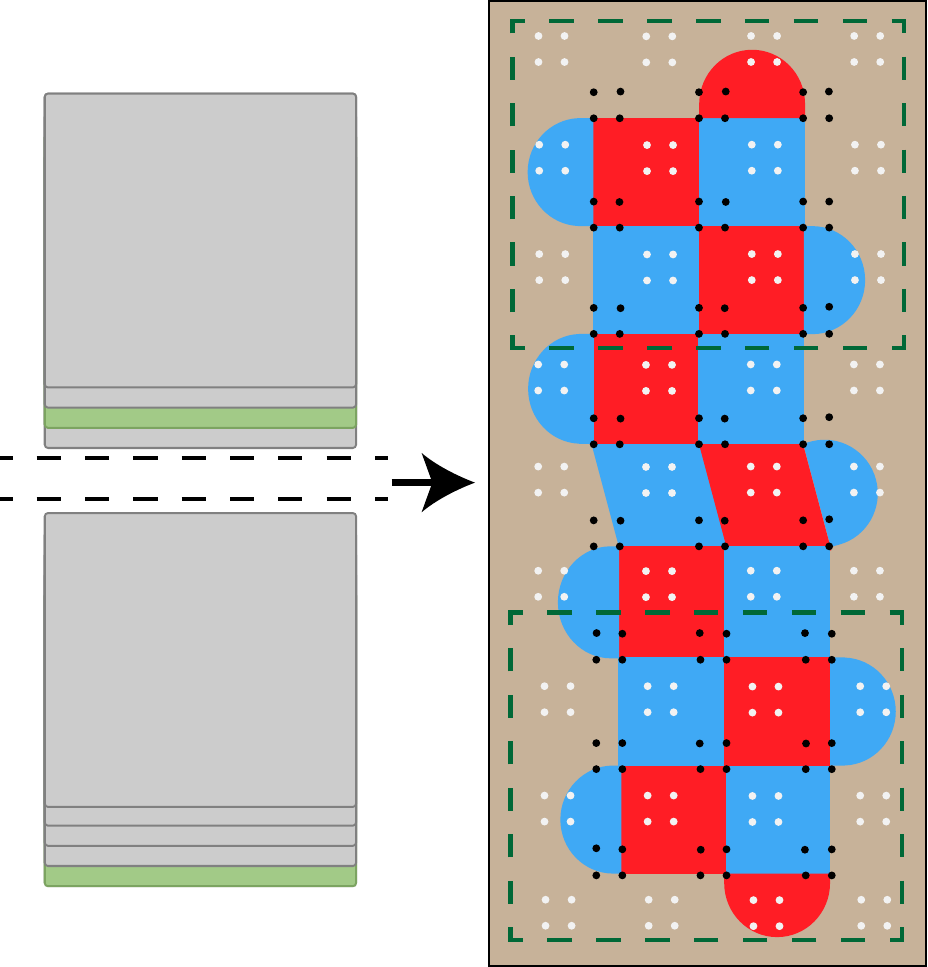}
    \caption{Performing a lattice surgery merge between two surface code qubits in different positions of separate interleaved groups. The key feature is the parallelogram-shaped stabilizers, which connect the bottom left logical qubit of the first group to the bottom right logical qubit of the second group. The stabilizers for the other logical qubits in each group are unchanged.}
    \label{fig:i_lattice}
\end{figure}

\begin{figure}
    \centering
    \includegraphics[width=0.75\linewidth]
    {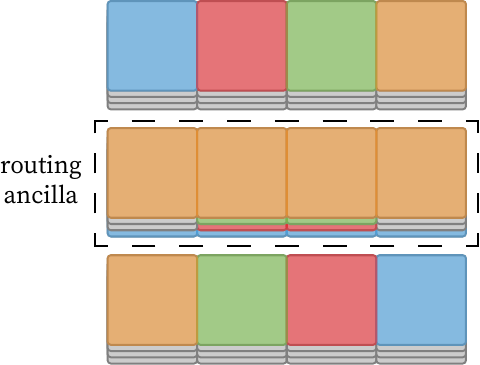}
    \caption{High-level example of routing four lattice surgery operations concurrently using groups containing four ancilla. Each group is an interleaving as shown in Figure~\ref{fig:multi-qubit}d, visualized as a stack.}
    \label{fig:i_routing}
\end{figure}

One option is to use lattice surgery techniques which require additional qubits to be used for routing space. Routing a lattice surgery operation between two logical qubits requires allocating a path of ancilla connecting the two qubits. These ancilla cannot be shared between concurrent operations, and so the routing problem is to identify non-intersecting paths in the ancilla space for a set of parallel operations. Programs with a high degree of parallelism will typically have high routing costs: either routing space is limited and parallel operations need to be serialized, increasing the program execution time, or additional routing space is allocated to support all parallel operations, increasing the ancilla qubit footprint. 

We propose a simple adaptation of lattice surgery to perform operations between interleaved groups, shown in Figure~\ref{fig:i_lattice}. Like standard lattice surgery, new stabilizers are measured between separate surface code patches, however, now the new stabilizers are also interleaved. For uninvolved qubits, we measure the old stabilizers as normal. But for qubits involved in an operation, we measure these new stabilizers, allowing us to choose how many lattice surgery operations to perform. Additionally, we can easily perform lattice surgery operations between qubits in differing positions of neighboring groups due to the Rydberg interaction radius. As a result, an all-to-all connectivity exists between qubits in adjacent interleaved groups. \paragraph{\textup{\textbf{Reduced Routing Costs:}}}
If we model routing ancilla as interleaved groups, this has notable consequences for the routing problem. By using interleaved groups of $k$ qubits, each ancilla group in our 2D array can now be used in $k$ operations, rather than just 1, augmenting the connectivity of the routing space. For a fixed number of routing qubits per program qubit, this improves the ability to support parallel operations. For example, Figure~\ref{fig:i_routing} shows an interleaved version of the compact block design~\cite{litinski2019game}, which requires $1.5n$ logical qubits for $n$ program qubits. With standard lattice surgery on the same number of routing ancilla, this example would require at least 2 time steps, whereas the interleaved version instead takes only 1 time step while preserving the ratio of ancilla qubits to program qubits. This also does not replace using transversal CNOT operations between program qubits in the same group, and so our circuit becomes a hybrid of transversal CNOTs and interleaved lattice surgery operations, depending on the location of the interacting qubits on our device.
\subsubsection{Atom Movement}
Another option is to physically move the atoms of a logical qubit so a transversal CNOT can be performed. Experimental demonstrations of mid-circuit movement of atoms have been achieved~\cite{bluvstein2022quantum}, and are a promising methodology for QC in atom arrays. 

This entails using acoustic-optic deflectors (AODs) which allows for the movement of a 2D array of trapping beams. This movement can be a translation across the device. For the purposes of a surface code, this means a single logical qubit's set of physical data qubits can all be moved in parallel. The 2D array is arranged over the data qubits and then the whole array is translated. 

This operation has a couple of architectural implications. It preserves the transversal CNOT discussed in Section~\ref{sec:single-iq}. As long as the time needed to move the atoms is small enough, this is also faster than a lattice surgery based approach. It also doesn't require routing ancilla. Notably, given that the whole movement array has to be translated together, we constrain ourselves to moving a single logical qubit at a time. Although future work could explore optimal methodologies to mitigate this movement constraint. 

Since the time taken for a movement-based approach scales as the device grows, we expect there to be a trade-off as long-range operations are needed. This trade-off is between the execution time and the reduced space cost from removing routing ancilla overhead. However, we anticipate both costs to be lower than lattice surgery for small programs. We run simulations quantifying this in Section~\ref{sec:compute_time}.

\section{Evaluation}\label{sec:evaluate}

One of the biggest hurdles for implementing QEC is the associated resource costs. Estimates for large-scale algorithms place execution times on the order of hours to months with tens of thousands to millions of physical qubits~\cite{gidney2021factor, beverland2022assessing}. To evaluate the efficacy of an interleaved architecture in reducing these resource costs, we run numerical simulations, since current devices are not able to support the benchmark circuits in this work. 

\subsection{Logical Error Rates of Primitive Operations}\label{sec:stim_results}

\begin{figure}
    \centering
    \includegraphics[width=\linewidth]{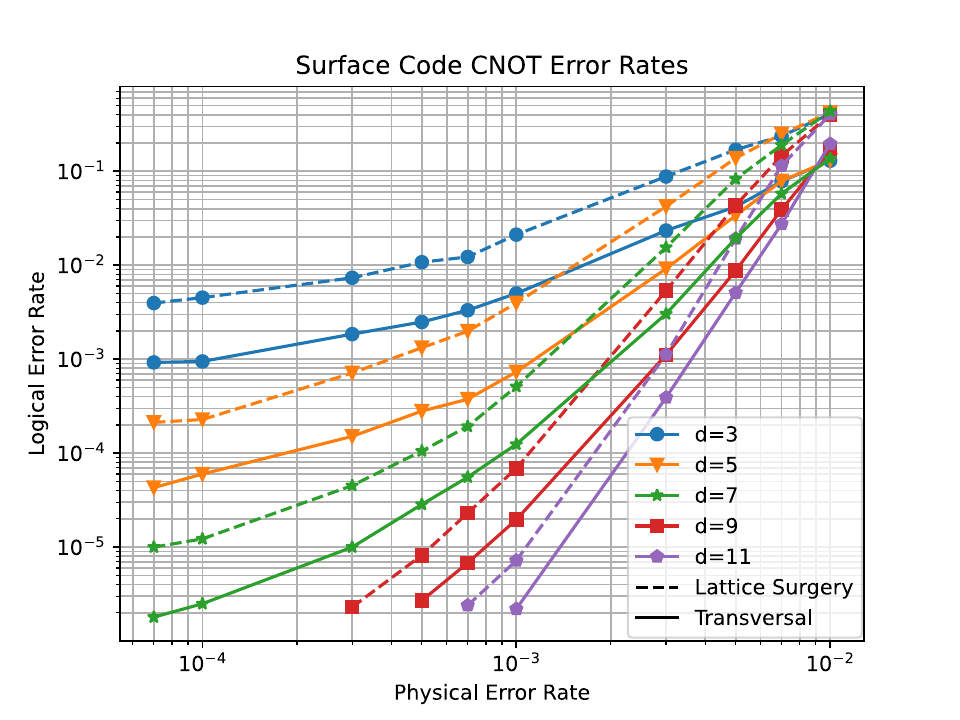}
    \caption{The logical error rate of a CNOT operation performed transversally in an interleaved architecture and via lattice surgery in a standard surface code architecture.}
    \label{fig:cnot-compare}
\end{figure}

\begin{figure}
    \centering
    \includegraphics[width=\linewidth]{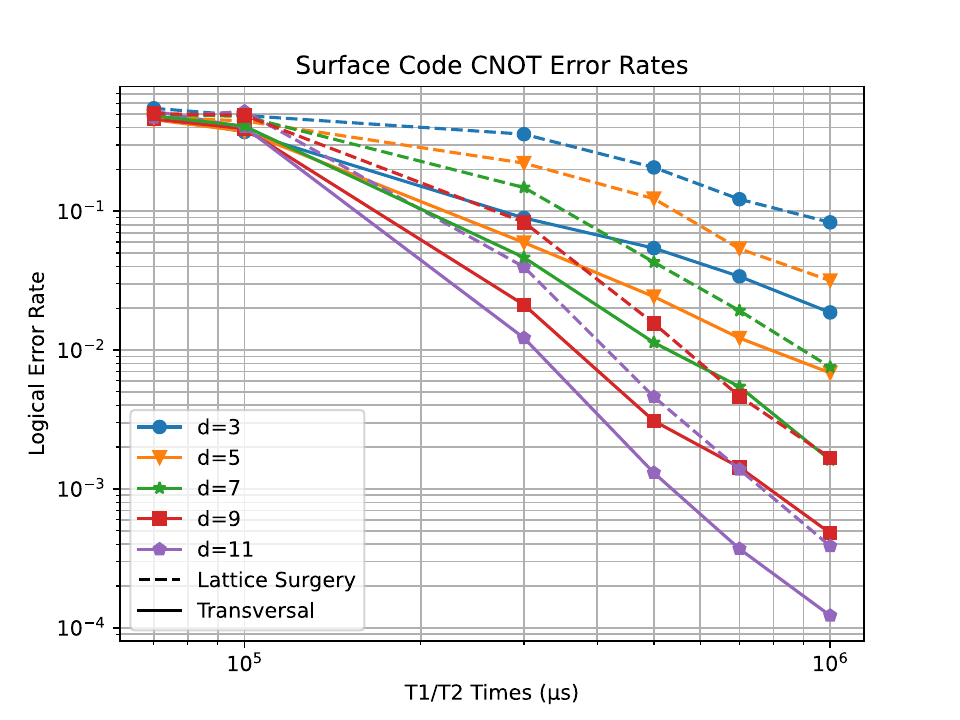}
    \caption{The logical error rate's sensitivity to coherence times for a CNOT performed transversally in an interleaved architecture and via lattice surgery in a standard surface code architecture.}
    \label{fig:cnot-compare-t1}
\end{figure}


We performed circuit-level simulations targeting Stim~\cite{gidney2021stim} and PyMatching~\cite{higgott2022pymatching} to estimate logical error rates of surface code QEC circuits. We describe logical circuits for a CNOT gate at the physical gate level and compile them into Stim's circuit format. We ran two simulations for a CNOT. In the first, we model an interleaved architecture and perform a transversal CNOT followed by $d$ stabilizer rounds. In the second, we model a standard surface code architecture and perform a lattice surgery CNOT as described in Figure~\ref{fig:lattice_cnot} which takes $2d$ stabilizer rounds. In all simulation experiments, stabilizer measurement circuits were modeled after Rydberg gate operations which were scheduled greedily. Figure~\ref{fig:cnot-compare} shows a typical threshold graph, demonstrating exponentially lower logical error rates for higher code distances at physical error rates below threshold. If we assume each CNOT's error rate, $p$, follows the simulated logical error rate, we can use this to estimate the necessary code distance to implement a target circuit. We observe that overall, a transversal CNOT in the interleaved architecture has lower logical error rates than a lattice surgery CNOT given the same code distance. For higher physical error rates, this separation can be a full code distance--a logical error rate that requires code distance $d$ in the standard surface code architecture only requires $d-2$ in the interleaved architecture. Since our number of physical qubits per logical qubits in the rotated surface code is $2d^2$, the difference in qubit footprint of the two schemes would be $2d^2 - 2(d-2)^2 = 8(d-1)$ for each program qubit. 

We also looked at the sensitivity of the logical error rate to coherence times as shown in Figure~\ref{fig:cnot-compare-t1}. Given that the transversal CNOT has longer stabilizer circuit depth, we expect it to be more sensitive to the qubit coherence time. Surprisingly, we don't observe this in our numerical simulations and instead find the logical error rate of the lattice surgery CNOT to decrease faster as coherence time is lowered. This is likely due to the longer logical circuit required for a lattice surgery CNOT which is 2d measurement cycles, compared to $d$ measurement cycles for the transversal CNOT. These results further confirm that the long coherence times of neutral atoms and the short operation time of the transversal CNOT mitigate any impacts on logical error rate due to the serialization of the stabilizer circuit.

\subsubsection{Error Modeling}

\begin{table}  
    \centering
    \begin{tabular}{ll}
        \multicolumn{2}{c}{Simulation Parameters}\\ 
        \hline
        Atom Spacing  & 10 µm \\ 
        Ancilla-Data Rydberg Radius & 28 µm \\
        Data-Data Rydberg Radius & 14 µm \\
        Movement Speed & 0.55 µm/µs \\\hline
        1Q Gate Time & 1 µs \\
        Rydberg Gate Time  & 5 µs \\
        Measurement Time  & 10 ms \\ 
        T1 Time & 1s \\
        T2 Time & 1s \\\hline
        1Q Gate Error & $0.1\%$\textbf{*} \\
        Rydberg Gate Error & $0.1\%$\textbf{*}  \\ 
        Measurement Error & $0.1\%$\textbf{*}
    \end{tabular}
    \caption{Parameters used in circuit-level simulations of the CNOT error rate based on recent experimental demonstrations. \textbf{*} We choose gate and measurement errors below the surface code threshold which is necessary to apply QEC.}
    \label{tab:sim_params}
\end{table}

There are two common ways to model errors in QEC simulations. The first is the phenomenological noise model. In this model, $X$/$Z$ errors occur on physical data qubits and measurement errors occur on physical ancilla qubits with independent probabilities $p$. This does not model the underlying circuit for measuring stabilizers and often estimates the surface code threshold error rate at $\sim2\%-3\%$. 

The second model is the circuit-level noise model which applies a depolarizing noise channel after all physical gates and adds measurement errors. The depolarizing channel equates to a Pauli error after each gate occurring with equal probability. Single qubit operations have $\{X,Y,Z\}$ applied with probability $\frac{p}{3}$ after each gate and two qubit operations have $\{I,X,Y,Z\}\otimes\{I,X,Y,Z\}\setminus II$ applied with probability $\frac{p}{15}$ after each gate. This requires a gate-level description of the stabilizer measurement circuit and results in lower estimates of $\sim0.5\%-1\%$\cite{higgott2023sparse} for the surface code threshold.

In our simulations we opt to use a circuit-level noise model. This fits a more detailed modeling of the stabilizer measurement circuit including serialized Rydberg gates in the interleaved architecture. We additionally model decoherence using popular Pauli twirling approximations~\cite{ghosh2012surface}. Table~\ref{tab:sim_params} contains the parameters we use in our simulations, which are based on recent experimental demonstrations~\cite{singh2022dual, bluvstein2022quantum}.

\subsection{Compute Time}\label{sec:compute_time}

\begin{figure}
    \centering
    \includegraphics[width=0.75\linewidth]{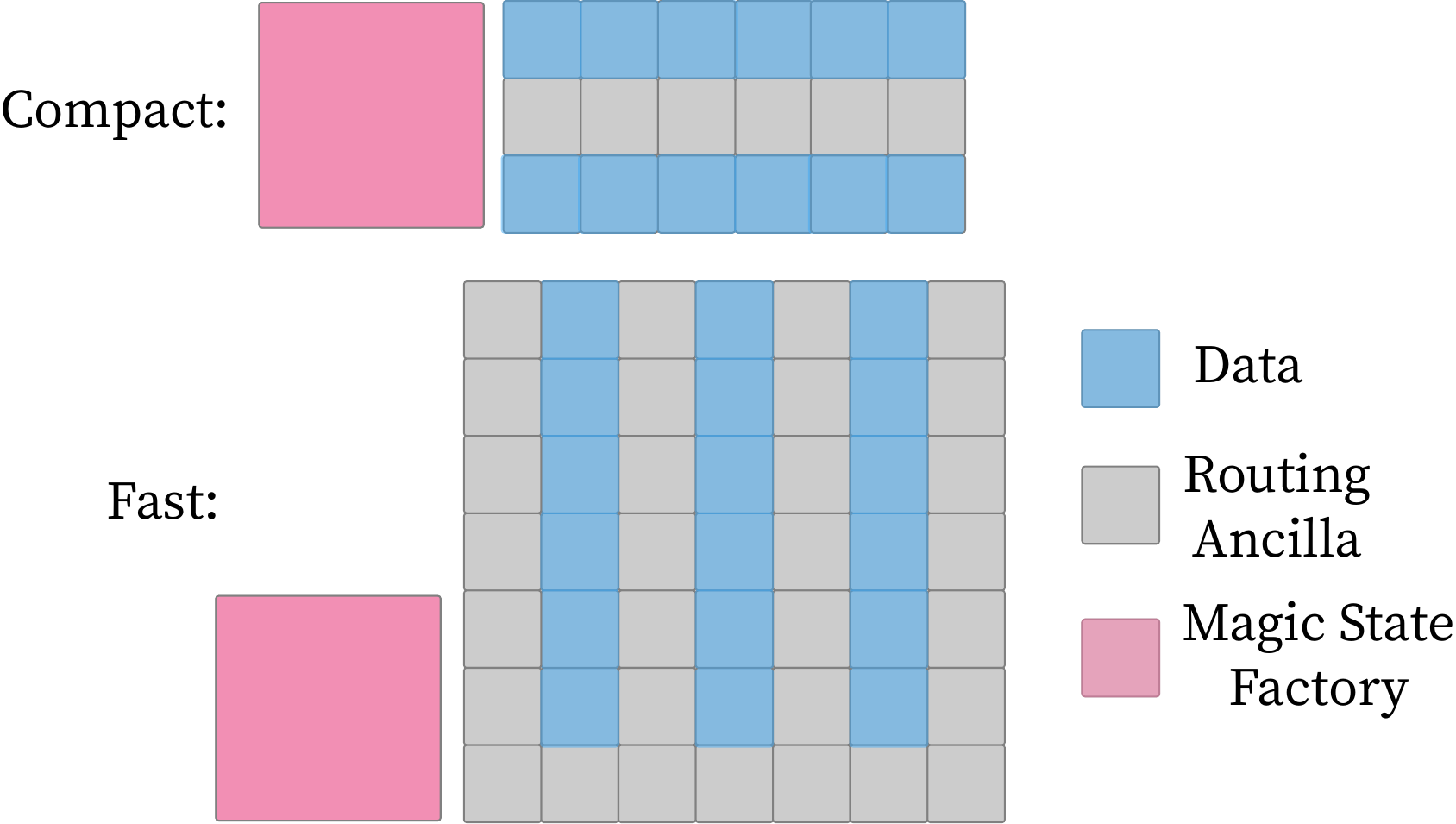}
    \caption{Surface code device layouts considered in estimating the overhead in total compute time from T and CNOT gates}.
    \label{fig:layouts}
\end{figure}

\begin{figure*}
    \centering
    \includegraphics[width=1\textwidth]{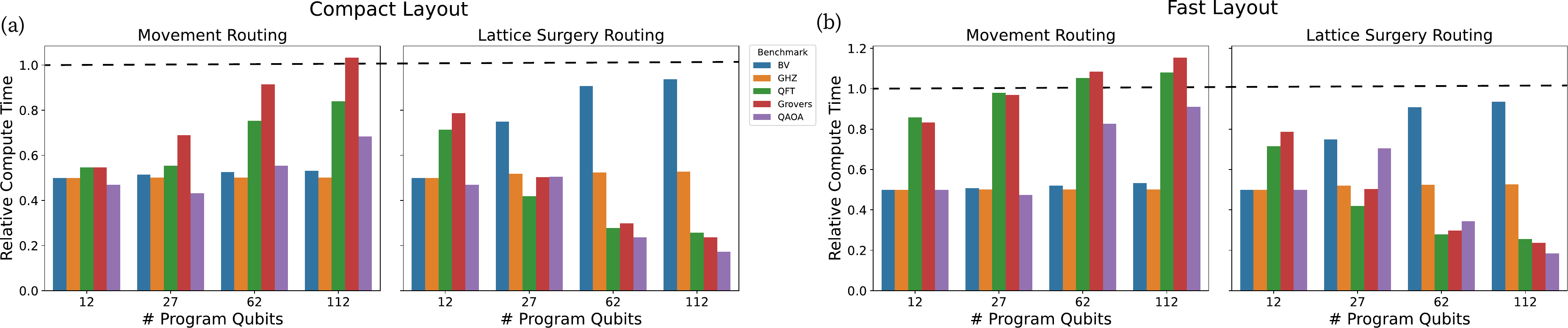}
    \caption{Estimated total compute time (lower is better) of benchmark programs using interleaved logical qubits and routing with movement or interleaved lattice surgery. Two different layouts from Figure~\ref{fig:layouts} are evaluated. Results are relative to a standard non-interleaved architecture that uses lattice surgery (dashed line). The simulations use interleaved groups of size 16 and the parameters in Table~\ref{tab:sim_params}.}
    \label{fig:compute-results}
\end{figure*}

To achieve universal quantum computation non-Clifford gates must be performed. These are not protected by the surface code, and so executing them requires different procedures than executing Clifford gates. A common non-Clifford gate to use for universality is the T gate. Protocols for using the T gate with the surface code require distilling of many noisy T gates into fewer T gates with lower error rates, a process called magic state distillation~\cite{bravyi2012magic}. These fewer T gates can then recursively be distilled until a target error rate is achieved. Proposals for large-scale systems using Clifford + T gates in the surface code require sections of the device, called magic state factories, be dedicated to performing magic state distillation. When executing the quantum program, magic states are then routed from factories to program qubits to perform T gates.

For larger-scale systems with 10s-100s of surface code qubits, routing multi-qubit operations, including T gates, across the device incurs a noticeable overhead. With lattice surgery, most estimates assume a qubit layout where the ratio of data qubits to routing qubits is fixed. Any variances in the routing costs are then observed in the overall compute time. To evaluate the performance of different approaches for scaling interleaved qubits, we opt to emulate this framework. We choose 2 qubit layout strategies~\cite{litinski2019game, beverland2022assessing} and estimate the total compute time from T and CNOT gates. The chosen layouts are shown in Figure~\ref{fig:layouts}. 

We estimate routing times by emulating a simplified surface code compiler flow. Benchmark circuits are synthesized into Clifford+T gates using Qiskit~\cite{Qiskit} and gridsynth~\cite{ross2014optimal}. The resulting circuits are scheduled using an as soon as possible policy and mapped using a greedy heuristic algorithm. Based on the results in Section~\ref{sec:stim_results}, our mapping heuristic prioritizes the ability to perform a transversal CNOT, mapping frequently interacting qubits to the same interleaved group. To route logical operations, we iterate through all T and CNOT gates in a given time slice and allocate the shortest path in the routing space for each gate. T gates are routed between the logical qubit and the magic state factory, and CNOT gates are routed between the two logical qubits. If a path in the routing space does not exist, for example due to other gates in the same time slice already occupying the routing space, then the gate is postponed a single time slice, delaying gates in future time slices. 
This methodology is not meant to be optimal, as compiling for lattice surgery is a hard problem~\cite{herr2017optimization} and an active area of research, but is meant to give a rough estimation of the relative routing costs of different approaches in the presence of varying levels of circuit parallelism.

We then estimate the total compute time of key gates: T and CNOT, that require routing across the device. In our estimations we use interleaved logical qubits with interleaved lattice surgery and with atom movement. The results are then calculated relative to a standard surface code embedding with lattice surgery.

\begin{figure*}
    \centering
    \hspace*{-6em}  
    \includegraphics[width=1.2\textwidth]{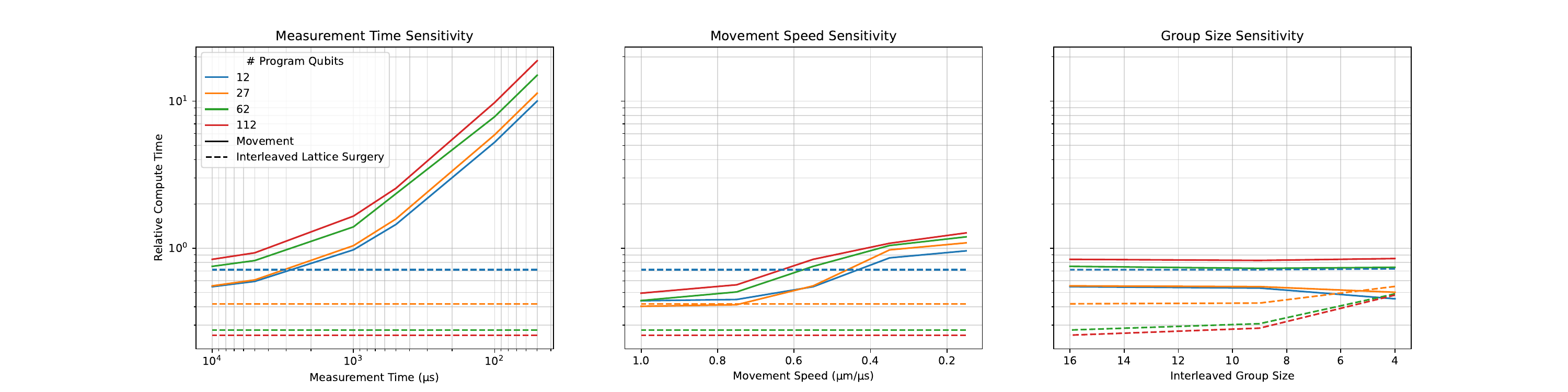}
    \caption{Compute time sensitivities (lower is better) to  different parameters of interleaved logical qubits. Results are relative to a standard non-interleaved architecture that uses lattice surgery. Simulations use the QFT benchmark and a compact layout. When not the focus of the sensitivity study, we fix interleaved group size at 16, measurement time at $10^4$ $\mu s$, and movement speed at $0.55$ $\mu m/\mu s$.}
    \label{fig:sensitivity}
\end{figure*}

\subsubsection{Benchmark Circuits}
For this work we choose a mixture of classic quantum benchmarks as well as scalable error correction circuits with varying amounts of parallelism. Our classic benchmarks include Bernstein-Vazirani, a near-term circuit with entirely serial gates but requires no synthesis (decomposition of non-Clifford gates) and QAOA (Quantum Alternating Operator Ansatz) a near-term variational algorithm which has parallelism dictated by the underlying graph structure as well as numerous small angle rotations. For these benchmarks we use the specification provided in the Supermarq benchmark suite~\cite{tomesh2022supermarq}. In the QEC regime, we explore the quantum Fourier transform (QFT) which is mostly serial but contains large numbers of small angles. We evaluate the 15-to-1 magic state distillation circuit~\cite{bravyi2012magic} which is a building block for distilling T gates. Finally, we consider Grover's algorithm, with a relatively small number of iterations, which searches simply for an integer meaning the entire circuit is a long sequence of multiply-controlled Toffoli gates which gets decomposed into large numbers of parallel CNOT and $T$ gates; no synthesis is necessary.

In these circuits, arbitrary rotations must be approximated with long sequences of $H$, $S$, and $T$ gates. This causes arbitrary rotations to become the bottleneck, as opposed to two qubit gates which are the bottleneck for hardware level NISQ circuits. Unless otherwise specified, all such approximations are assumed to be within $\epsilon \le 1e-10$. 


\subsubsection{Compute Time Results}

Figure~\ref{fig:compute-results} shows the results of our compute time estimates relative to standard lattice surgery for fast and compact device layouts. Routing via atom movement uses a transversal CNOT and $d$ measurement rounds, while a lattice surgery CNOT takes $2d$ measurement rounds. Not shown in this figure is the results of the 15-to-1 magic state distillation circuit. We evaluate the 16 qubit version of this circuit which fits in a single interleaved group of size 16. The relative compute time is 0.41 and 0.47 for the compact and fast layouts, respectively. Since the circuit fits in a single interleaved group, routing was not necessary.

For the rest of the benchmark circuits we find that small scale ($<\sim50$ qubits), Bernstein-Vazirani (BV), and GHZ programs benefit the most from a movement based routing scheme between interleaved groups. This is in agreement with our expectation. For small circuits, routing distances are short and so the time spent moving a logical qubit and performing a transversal CNOT is still less than performing a lattice surgery CNOT. For the BV and GHZ programs, the programs are entirely serial, meaning the increased parallelism from interleaved lattice surgery gives no benefit. Additionally, program qubits only interact with few other qubits and the programs do not contain T gates, causing routing distances to be short. However, larger programs ($>\sim50$ qubits) that include T gates (QFT, Grovers, and QAOA) require lots of routing across the device between logical qubits and magic state factories. These programs also exhibit circuit-level parallelism. As a result, routing serially via atom movement becomes slower compared to interleaved lattice surgery.

We also note the difference in performance between the two layouts. Using the compact layout gives the largest improvement in relative compute time. The key takeaway is interleaved lattice surgery and atom movement both perform well in the compact layout despite it having less routing ancilla. This is in contrast to the baseline standard architecture. Due to less routing space, the baseline has a higher frequency of routing collisions. This requires delaying subsequent operations until the routing ancilla is available again after $2d$ measurement rounds.

In summary, we believe our results motivate the use of the compact layout when using interleaved groups, requiring the least amount of routing ancilla. Additionally, for programs in the QEC regime routing should use hybrid operations, where short-range interactions are routed using atom movement and parallel, long-range interactions are routing using interleaved lattice surgery. 



\subsubsection{Sensitivity Results}

We perform three sensitivity studies with the goal of understanding the impact of device-level features as devices scale to support larger programs. We also note their impact on the program size cross-over point between atom movement and interleaved lattice surgery.

The time it takes to perform a measurement operation has the largest impact on the performance of routing via atom movement. If devices improve and lower the physical measurement time, the time needed to move atoms becomes a larger hindrance. This would shift the cross-over to smaller program sizes. Alternatively, if the speed at which atoms can be safely moved is improved then the cross-over point shifts towards larger program sizes.

The achievable interleaved group size is related to how high a Rydberg state atoms can be excited to, which defines the ancilla-data interaction range. Larger group sizes enable better performance using interleaved lattice surgery for larger programs due to the higher capacity routing space. However, we don't observe much benefit increasing the group size past 9.

\section{Conclusions}\label{sec:conclusions}

As quantum computers move away from the NISQ era and towards quantum error correction (QEC), co-designing architectures with the underlying hardware will be crucial in accelerating the realization of impactful quantum algorithms. In this work, we proposed a novel surface code architecture using groups of interleaved logical qubits on quantum computers made from 2D arrays of neutral atoms. Our architecture enables a transversal CNOT with all-to-all connectivity within groups of logical qubits. Compared to a lattice surgery CNOT we find the transversal approach has a lower logical error rate and requires no logical ancilla, reducing the cost of near-term QEC demonstrations.

We also look at the impact of the architecture for larger scale devices in the future. We propose a tiling of many groups of interleaved qubits and analyze two methodologies for multi-qubit operations between groups: interleaved lattice surgery and atom movement. In the regime of programs < 50 qubits, we find atom movement leads to to the shortest execution times. For programs > 50 qubits, interleaved lattice surgery performs best due to the need for long-range operations. In both regimes, the best methodology demonstrates a 2x-5x speedup in compute time compared to a standard lattice surgery architecture, motivating the use of a hybrid design including both for large-scale devices.

Given these results, we hope this work motivates future exploration of neutral atom devices for quantum error correction applications, as well as further co-design of QEC architectures with the underlying hardware.



\section{Acknowledgements}
This work is funded in part by EPiQC, an NSF Expedition in Computing, under award CCF-1730449; in part by STAQ under award NSF Phy-1818914; in part by the US Department of Energy Office of Advanced Scientific Computing Research, Accelerated 
Research for Quantum Computing Program; and in part by the NSF Quantum Leap Challenge Institute for Hybrid Quantum Architectures and Networks (NSF Award 2016136), in part based upon work supported by the U.S. Department of Energy, Office of Science, National Quantum 
Information Science Research Centers, and in part by the Army Research Office under Grant Number W911NF-23-1-0077. The views and conclusions contained in this document are those of the authors and should not be interpreted as representing the official policies, either expressed or implied, of the U.S. Government. The U.S. Government is authorized to reproduce and distribute reprints for Government purposes notwithstanding any copyright notation herein.
FTC is Chief Scientist for Quantum Software at Infleqtion and an advisor to Quantum Circuits, Inc.
\vspace{2em}


\bibliographystyle{plain}
\bibliography{refs}

\end{document}